\documentclass[english]{article}
\usepackage[T1]{fontenc}
\usepackage[utf8]{inputenc}
\usepackage{lmodern}
\usepackage{color}
\usepackage{bm}
\usepackage{amsmath}
\usepackage{amssymb}
\usepackage{graphicx}
\usepackage{subcaption}
\usepackage{authblk}
\usepackage{placeins}
\usepackage{color}
\usepackage{bm}
\usepackage{amsmath}
\usepackage{amssymb}
\usepackage{booktabs}
\usepackage{siunitx}
\usepackage{float}
\usepackage{array}
\usepackage{xspace}
\usepackage{tabularx}

\usepackage{epsfig}%\usepackage{subfigure}
\usepackage{amsfonts}\usepackage{array}
\usepackage{theorem}%\pagestyle{myheadings}

\newcommand{\br}{{\bf r}\xspace}
\newcommand{\bd}{{\bf d}\xspace}
\newcommand{\mbr}{\ensuremath{\mathbf{r}}}
\newcommand{\mbd}{\ensuremath{\mathbf{d}}}

\newcommand{\eg}{{\textit{e.g.}}\xspace}

\newcommand{\bml}{\begin{subequations}}
\newcommand{\eml}{\end{subequations}}
\allowdisplaybreaks[1]

\newcommand{\qed}{\nobreak \ifvmode \relax \else
      \ifdim\lastskip<1.5em \hskip-\lastskip
      \hskip1.5em plus0em minus0.5em \fi \nobreak
      \vrule height0.75em width0.5em depth0.25em\fi}
\usepackage{vmargin}
\setpapersize{custom}{8.5in}{11.0in}
\setmarginsrb{1.0in}{1.0in}{1.0in}{1.0in}%
            {0pt}{10mm}{0pt}{0mm}

\let\originalleft\left
\let\originalright\right
\renewcommand{\left}{\mathopen{}\mathclose\bgroup\originalleft}
\renewcommand{\right}{\aftergroup\egroup\originalright}

\usepackage{mathtools}

\usepackage{multirow}
\usepackage{siunitx}
\makeatother
\usepackage{babel}
\usepackage{caption}
\usepackage{dcolumn}
\newcolumntype{d}[1]{D{.}{.}{#1}}
\usepackage{url}
\usepackage{hyperref}
\usepackage[
  backend=biber,
  style=ieee,
  sorting=none,      % Sort references by citation order
  citestyle=numeric-comp,  % Ensure citations are numeric
  maxnames=3,       % Display all authors
  url=true,          % Enable URL handling
  doi=true,          % Enable DOI handling
]{biblatex}

\begin{document}

\title{Boosting the Efficiency of the Differential Algebra-based Fast Multipole Method Using Symbolic Differential Algebra}

\author[1]{He Zhang}
\affil[1]{Thomas Jefferson National Accelerator Facility, Newport News, VA, USA}

\date{}% if you don’t want a date

\maketitle

\begin{abstract}
	The Fast Multipole Method (FMM) computes pairwise interactions between particles with an efficiency that scales linearly with the number of particles. The method works by grouping particles based on their spatial distribution and approximating interactions with distant regions through series expansions. Differential Algebra (DA), also known as Truncated Power Series Algebra (TPSA), computes the Taylor expansion of a function at a given point and allows users to manipulate Taylor expansions as easily as numerical values in computation. This makes it a convenient and powerful tool for constructing expansions in FMM. However, DA-based FMM operators typically suffer from lower efficiency compared to implementations based on other mathematical frameworks, such as Cartesian tensors or spherical harmonics.	
	To address this, we developed a C++ library for symbolic DA computation, enabling the derivation of explicit expressions for DA-based FMM operators. These symbolic expressions are then used to generate highly optimized code that eliminates the redundant computations inherent in numerical DA packages. For individual FMM operators, this approach achieves a speedup of 20- to 50-fold.	
	We further evaluate the numerical performance of the enhanced DA-FMM and benchmark it against two state-of-the-art FMM implementations, pyfmmlib and the traceless Cartesian tensor-based FMM, for the Coulomb potential. For relative errors on the order of $10^{-7}$ or higher, the enhanced DA-FMM consistently outperforms both alternatives.
\end{abstract}

\section{Introduction}
The fast multi-particle method (FMM) is a fast and accurate
method for computing pairwise particle interactions \cite{greengard87}. The FMM scales linearly with the number of particles. This holds for any non-oscillating interaction, commonly referred to as the kernel function. It has been
widely used in many areas, such as
astronomy \cite{warren1992astrophysical}, electrical
engineering \cite{gumerov2004fast}, molecular
dynamics \cite{board1992accelerated}, fluid
dynamics \cite{salmon1994fast}, and crack detection \cite{nishimura1999fast}. The strategy of FMM can be briefly described as follow. First, we divide the whole domain under study into boxes recursively, ensuring that each box contains approximately the same number of particles. If the particles are nearly uniformly distributed, the entire space can be divided into equally sized boxes. The hierarchical relationship between the boxes can be represented as a full tree. If the particles are far from uniformly distributed, finer boxes will be generated where the particles are more densely packed while coarser boxes will be generated where the particles are more sparsely packed, which leads to a partially overlapping tree of the boxes. Then we approximate the interactions between the boxes using expansions. For each box, its contribution to its far region is calculated as the multipole expansion, and the contribution inside the box from its far region is calculated as the local expansion. The interaction inside the near region of the box is calculated directly using the pairwise formula. FMM expansions can be constructed using different mathematical methods. As long as the following seven formulas are available, an FMM can be built: (1) calculate the multipole expansion of a box  from the particles inside (P2M), (2) calculate the multipole expansion of a box by adding up the multipole expansions of its child boxes (M2M), (3) convert a multipole expansion in a box into a local expansion in another box in the far region (M2L), (4) calculate the local expansion of a box from the particles in a  far region box (P2L), (5) calculate the local expansion of a box from the local expansions of its parent box (L2L), (6) calculate the interaction on the particles from a local expansion (L2P), and (7) calculate the interaction on the particles from a multipole expansion (M2P). 

Differential algebra (DA), which is also referred to as truncated power series algebra (TPSA), was invented in late  1980s to study the dynamics of particle beams in a repeated system \cite{pada}. It provides a convenient way to manipulate Taylor expansions of functions at arbitrary points. Using computer libraries that support DA calculation, we can easily calculate the FMM expansions for a given kernel. The DA-based FMM for the Coulomb kernel was developed in 2010 \cite{FMMCPO2010} and algorithms and codes based on it have been developed for particle-based beam dynamic simulations with strong interactions between the particles, \eg,  the space charge effect and the electron cooling effect \cite{tao2011quantitative,tao2012space,portman2013computational,zhang2015differential,abeyratne2019adaptive,al2021efficient,tencate2021differential}. However, comparing with FMMs using other mathematical techniques, the DA-based FMM suffers from a lower efficiency. To improve the efficiency of the DA-based FMM, we developed a new library that carries out symbolic DA calculations. Using the symbolic DA library, we can obtain the explicit expressions of all the DA-based FMM expansions. These explicit expressions eliminate redundant steps in time-consuming DA calculations and enable us to write more efficient code by  strategically arranging computations - rather than directly calling the numerical DA libraries. Using the Coulomb kernel as an example, we demonstrated that the computational cost can be significantly reduced for all the FMM operators using the code generated from the symbolic DA expressions.The enhanced DA-FMM matches or surpasses the efficiency of state-of-the-art FMM implementations.

\section{DA and DA-based FMM for Coulomb Kernel}\label{sec:da}

A new data type has been introduced in the libraries that support DA calculation. In the following, we will call it DA vector. A DA vector saves a sequence of numbers, each of which is the coefficient of the corresponding unique monomial of the pre-selected bases. Hence the DA vector represents a polynomial of the bases. The DA vector provides a natural way to calculate and save the Taylor expansion of a function at a given point. Libraries that support DA calculations are available in several popular programs for particle accelerator design and simulations, such as COSY Infinity \cite{COSYCAP04}, MAD-X \cite{grote2003mad}, and PTC \cite{forest2002introduction}. Stand-alone DA/TPSA libraries include DACE \cite{massari2018differential, DACE} and cppTPSA/pyTPSA \cite{zhang2024cpptpsa}. With these libraries, DA vectors can be used as numbers in the calculation. Fundamental math operators and common functions for DA vectors are provided by the library. The composition of two functions, represented as DA vectors, are usually supported, too. We give a very brief introduction on DA/TPSA from a practical computational perspective in Appendix~\ref{ap:da}. Please refer to \cite{chao2022special} for a more detailed instruction or \cite{AIEP108book} for the complete theory.

To construct the FMM expansions in DA, we need to find proper small variables as the bases for the DA vectors. After
that, most computations can be carried out automatically by the DA library. In the following, we list the formulas
for the DA-based FMM operators for the Coulomb kernel. 

\subsection{P2M and M2P}
For $n$ particles with charges $q_i$ and positions $\mathbf{r}_i$, the electrostatic potential $\phi$
at a far away position \br can be expressed as in Eq.~(\ref{eq:P2M_phi}).
\begin{equation}
  \phi = \sum_{i=1}^n \frac{q_i}{|\mathbf{r}_i - \mathbf{r}|} = d \cdot \phi_{\mathrm{M}}
  \label{eq:P2M_phi}
\end{equation}
where $\mathbf{d} = \mathbf{r}/r^2$ and 
\begin{equation}
  \phi_{\mathrm{M}} = \sum_{i=1}^n \frac{q_i}{\sqrt{1+r_i^2d^2+2\mathbf{r}\cdot \mbd}}.
  \label{eq:P2M}
\end{equation}
We choose the small variable \bd  as the bases for the DA vectors. $\phi_{\mathrm{M}}$ is the multipole expansion, which is represented as the Taylor expansion with respect to \bd. The calculation of    $\phi_{\mathrm{M}}$ can be carried out by a DA library.

\subsection{M2M}
To calculate the multipole expansion  of a parent box, we translate the multipole expansions of all the child boxes to the center of the parent box and take the summation. Assuming the center of the child box is $\mathbf{O}(0,0,0)$ and the center of the parent box is $\mbr_0(x_0,y_0,z_0)$, the new DA bases in the parent box frame is $\mathbf{d}^\prime = \mathbf{r}^\prime/r^{\prime 2}$ where $\mathbf{r}^\prime = \mathbf{r} - \mathbf{r}_0$. The relation between the old bases and the new bases can be represented as a map $M_1: \mathbf{d} = (\mathbf{d}^\prime + d^{\prime 2}\mathbf{r}_0) \cdot R$, where $R = 1/(1+r_0^2d^{\prime 2} + 2\mathbf{r}_0\cdot \mathbf{d}^\prime)$. The norms of the bases in the two frames satisfy $d = d^\prime \cdot \sqrt{R}$. The electrostatic potential $\phi^\prime$ in the parent box frame can then be calculated as the composition of the potential $\phi$ in the child box frame with the transfer map $M_1$, as shown in Eq.~(\ref{eq:m2m_phi}).
\begin{equation}
  \phi^\prime = \phi \circ M_1 = d^\prime \cdot \phi_{\mathrm{M}}^\prime.
  \label{eq:m2m_phi}
\end{equation}
$\phi_{\mathrm{M}}^\prime$ is the multipole expansion in the parent box and
\begin{equation}
  \phi_{\mathrm{M}}^\prime = \sqrt{R}\cdot(\phi_\mathrm{M} \circ M_1).
  \label{eq:m2m}
\end{equation}

\subsection{M2L}
When two boxes are well separated, we can convert the multiple expansion from the source box into a local expansion at the object box. Asumming the source box centered at $\mathrm{O}(0,0,0)$ and the object box centered at $\mbr_0 (x_0,y_0,z_0)$, we choose the DA bases in the object box frame as $ \mathbf{d}^\prime = \mathbf{r}^\prime = \mathbf{r}-\mathbf{r}_0 $. In the object box, \br is close to $\mbr_0$ and hence $\mbd^\prime$ is a small variable. The relation between the old bases and the new bases can be represented as a map $ M_2 : d = (\mathbf{r}_0 + \mathbf{r}^\prime) \cdot R $ with $ R = 1/ | \mathbf{r}_0 + \mathbf{r}^\prime|$. The electrostatic potential in the object box can be represented as a local expansion $\phi_\mathrm{L}$ as in Eq.~(\ref{eq:m2l}).
\begin{equation}
  \phi_{\mathrm{L}} = \phi \circ M_2 = (d \cdot \phi_{\mathrm{M}}) \circ M_2 = \sqrt{R} \cdot (\phi_{\mathrm{M}} \circ M_2).
  \label{eq:m2l}
\end{equation}
Here we used the result that $d$ in the source box frame will be converted into $\sqrt{R}$ in the object box frame.

\subsection{P2L and L2P}
For $n$ particles with charges $q_i$ and positions $\mathbf{r}_i$ outside the near region of an object box centered at $\mbr_0 (x_0,y_0,z_0)$, the electrostatic potential due to these particles inside the object box can be represented as a local expansion $\phi_{\mathrm{L}}$ as in Eq.~(\ref{eq:P2L}),
\begin{equation}
  \phi_{\mathrm{L}} =\sum_{i=1}^n \frac{q_i}{|\mathbf{r} - \mathbf{r}_i|} = \sum_{i=1}^n \frac{q_i}{|\mathbf{r}_0-\mathbf{r}_i+\mathbf{d}^\prime|},
  \label{eq:P2L}
\end{equation}
when we choose $\mathbf{d}^\prime = \mathbf{r}-\mathbf{r}_0= \mathbf{r}^\prime$ as the DA bases.
\subsection{L2L}
The local expansion in a parent box can be translated to its child boxes. Assuming the center of the parent box is $\mathbf{O}(0,0,0)$ and the center of the child box is $\mbr_0(x_0,y_0,z_0)$, the new DA bases in the child box is  $ \mathbf{d}^\prime= \mathbf{r} - \mathbf{r}_0$. The transfer map $M_3$ between the old and the new bases is simply a linear shift: $M_3: \mathbf{d} = \mathbf{r}_0 + \mathbf{d}^\prime$. The local expansion in the child box can be calculated using Eq.~(\ref{eq:l2l}).
\begin{equation}
  \phi_\mathrm{L}^\prime = \phi_\mathrm{L} \circ M_3.
  \label{eq:l2l}
\end{equation}

\section{Boosting the efficiency of the FMM operators using symbolic DA}
\begin{table}[b]
	\centering
	\caption{Numerical DA for $f(x,y,z)=1/\sqrt{x^2+y^2+z^2}$ at point (1,1,1) and symbolic DA for the same function up to the second order}
	\label{tab:DA_SDA}
	\begin{tabular}{rlc}
		\toprule
		\multicolumn{1}{c}{DA} & \multicolumn{1}{c}{SDA} & \multicolumn{1}{c}{Orders} \\
		%		DA & SDA & Orders \\
		\hline
		5.773502691896258e-01   &  $1/r$   & 0 0 0   \\
		-1.924500897298753e-01   &  $-x/r^3$   & 1 0 0  \\
		-1.924500897298753e-01   &  $-y/r^3$   & 0 1 0  \\
		-1.924500897298753e-01   &  $-z/r^3$   & 0 0 1  \\
		8.012344526598184e-18   &  $1.5x^2/r^5 - 0.5/r^3$   & 2 0 0  \\
		1.924500897298753e-01   &  $3xy/r^5$  & 1 1 0  \\
		1.924500897298753e-01   &  $3xz/r^5$   & 1 0 1  \\
		8.012344526598184e-18   &  $1.5y^2/r^5 - 0.5/r^3 $  & 0 2 0  \\
		1.924500897298753e-01   &  $3yz/r^5 $  & 0 1 1  \\
		8.012344526598184e-18   &  $1.5z^2/r^5 - 0.5/r^3$  & 0 0 2  \\
		\bottomrule
	\end{tabular}
\end{table}
Although DA provides a convenient way to construct the FMM operators, we have found the DA-based operators have a relatively low efficiency when compared with its peers, \eg the Cartesian tensor-based FMM \cite{shanker07,huang2018improve} or the spherical harmonic-based FMM \cite{greengard87,greengard2002new}. One possible reason is the time-consuming DA operators involved in the computation. For example, the inverse of a DA vector needs to be calculated in a order-by-order manner. For a $n$-th order DA vector, we need to iterate $n$ times to obtain the inverse of the same order. In FMM, the iteration has to be repeated on all the particles when we calculate the P2M operator, which could result in a lower efficiency. If we know the explicit expression of the P2M operator, we can omit the intermediate steps to save time. In addition, the DA libraries have to be compatible with arbitrary dimension and arbitrary order and may need complicated data types and algorithms to achieve that, while we usually deal with fixed dimension and order in FMM. The overhead we paid for the generality may also lower the efficiency of the code. Knowing the explicit expressions of the operators, we can develop codes that focus on the computation  using simple data types and algorithms for better performance. However, the numerical DA cannot give us the explicit expressions. We need a DA library that support symbolic calculation. 

Based on the numerical DA package, cppTPSA \cite{zhang2024cpptpsa}, we developed a symbolic DA (SDA) package \cite{SDAlink}. The SDA vector and the operators are defined in the exactly same way as in DA. All the DA calculations are carried out using exactly the same algorithms in cppTPSA, except they are implemented on symbols by employing the SymEngine library \cite{Fernando2024SymEngine}, rather than numbers. The key difference between DA and SDA is this: A DA vector is evaluated at a specific point of a function, resulting in numerical coefficients. Conversely, an SDA vector contains expressions of the function's variables - these are defined as symbols and hold true for all valid values of those variables. Table~\ref{tab:DA_SDA} shows an example of a second order DA vector of $f(x,y,z)=1/\sqrt{x^2+y^2+z^2}$ at point (1,1,1) and an SDA vector for the same function at the same order. The first column shows the coefficients in the DA vector, the second column shows the coefficients in the SDA order, and the third column show the orders of the bases. To simplify the expressions in the SDA result, we introduced the fourth variable $r$ and let $r=\sqrt{x^2+y^2+z^2}$. The change of variables can be  carried out easily in the following two ways. First, we can expand the squares in the expression of $f=1/\sqrt{(x+d_1)^2 + (y+d_2)^2 + (z+d_3)^2}$ and we will obtain $f = 1/\sqrt{r^2+2xd_1+2yd_2+2zd_3+d_1^2+d_2^2+d_3^2}$ and $r$ is now naturally involved. Second, instead of calculating the whole function in DA, we calculate $v = (x+d_1)^2 + (y+d_2)^2 + (z+d_3)^2$ first. The constant part, the first term with orders (0,0,0), of $v$ is $x^2 + y^2 + z^2$. Reset the value of the constant part by $r^2$ and then carry out the square root and the inverse. We will obtain the expressions in table~\ref{tab:DA_SDA} using either method. The second method shows that we do not depend on analytical expressions. Any complicated expression in an intermediate step can be replaced by a variable and it will be inherited in the following calculation to simply the final result. If we substitute (1,1,1) for $(x,y,z)$ in the expressions, we will get the same values for the coefficients as shown in the first column. Obviously, the expressions holds for other values of $(x,y,z)$ except for (0,0,0). Using the SDA library, we can obtain the explicit expressions of all the DA-based FMM operators and develop codes for high efficiency calculation.

\FloatBarrier
In the following, when an operator is computed by the code based on the expressions obtained from respective SDA calculations, we will call it an SDA operator for simplicity. When an operator is computed by calling the numerical DA library, we will call it a DA operator. 
For each FMM operator, we obtain explicit expressions for all coefficients using SDA up to order 10. A Python-based parser and code generator, developed with assistance from ChatGPT~\cite{OpenAIChatGPT}, analyzes these symbolic expressions and automatically generates the corresponding C++ functions.The parser performs two main functions: 1. It uses regular expressions to analyze the SDA output and extract all terms, including the sign (positive or negative), numerical coefficient, base variables, and the corresponding orders of those bases. 2.It identifies repeated terms to eliminate redundant computations. With this information, the code generator produces C++ code to perform the required calculations efficiently.  To verify the correctness of the generated code, we generate 10,000 random instances, use the C++ implementation to compute the results, and compare them order by order with those obtained from the numerical DA code. For each order, we calculate the error between the coefficients of each term as $\delta = C_\mathrm{sda}-C_\mathrm{da}$ and the relative error as $\delta_\mathrm{r} = \delta/C_\mathrm{da}$, where $C_\mathrm{sda}$ refers to the coefficient of the SDA operator and $C_\mathrm{da}$ the coefficient of the DA operator. Among all coefficients of the same order, we record the one with the largest absolute error  $|\delta|$ and the corresponding  absolute relative error $|\delta_\mathrm{r}|$. For example, to verify the SDA P2M operator, we generate 10,000 random particles uniformly distributed inside a cube box centered at the origin ranging from -1 to 1 in each direction, compute the multipole expansions using the SDA P2M operator, and benchmark the results against direct DA P2M computations. To verify the SDA M2M operator, we apply it to convert those multipole expansions into local expansions inside a randomly positioned box, separated by one intermediate box from the box located at the origin. The results are compared against those produced by the DA M2M operator. All other SDA operators are verified using similar procedures. All SDA FMM operators are also benchmarked for computational efficiency against their DA counterparts by measuring and comparing the average execution time over the 10,000 test instances. We will present the results in the following subsections. All numerical experiments are conducted on a desktop PC equipped with an Intel i5-14600 CPU and 16~GB of memory.

\subsection{P2M and P2L}

From a computational and programming standpoint, the P2M and P2L operators exhibit highly similar structures. Both accept three floating-point inputs corresponding to the spatial coordinates of a source charge and return a full-rank DA vector representing the multipole/local expansion of the potential. The underlying computations in both cases involve a square root and an inverse operation. Deriving explicit expressions for the P2M and P2L operators is straightforward. We define four symbolic variables, $x$, $y$, $z$, and $r$, with  $r = \sqrt{x^2+y^2+z^2}$. For P2M, $(x,y,z)$  denotes the position of the charge relative to the center of the source box.  Using SDA to carry out the calculation of Eq.~(\ref{eq:P2M}), all the coefficients of the results are expressed as polynomials in $x, y, z$ and $r$.  In Table~\ref{tab:cm}, we list the order in column 1, the value of the coefficient in column 2, the absolute value of $\delta$ in column 3, and the absolute value of the corresponding $\delta_\mathrm{r}$ in column 4. The constant part and all the first order coefficients agree exactly. Errors appear starting from the second order.  As the order goes up from two to ten, we observe that the maximum error increased by five orders from $4.44\times 10^{-16}$ to $2.62\times 10^{-10}$. Meanwhile, the absolute value of the coefficients also increased by four orders and the relative errors remain low. Except for the fifth order with $\delta_\mathrm{r}=3.69\times 10^{-14}$, the corresponding relative errors for all the other orders are below $1\times 10^{-14}$. We conclude the SDA P2M operator agree well with the DA P2M operator. Then we benchmark the efficiency of the SDA P2M against DA P2M with the cut-off order $p$ from two to ten. In Table~\ref{tab:cm}, we list the execution time for DA P2M in column 5, the execution time for SDA P2M in column 6, and the relative runtime of SDA P2M compared to DA P2M, expressed as a percentage, in column 7. As expected, higher-order P2M operators take longer to evaluate than lower-order ones because they contain more terms. When the expansion order is fixed, the execution time for an SDA P2M operator is significantly shorter than that of its DA counterpart: it remains below 8\% across all cases and drops below 3\% when the order exceeds five.

When a parent box locates in the near region of a childless box, P2L operator calculates the local expansion of the potential inside the target parent box due to the charges inside the childless box. The expression of the P2M operator can be obtained as  polynomials in $x, y, z$ and $1/r$ and $x, y, z$ denotes the position of the charge relative to the center of the target box. In Table~\ref{tab:cl}, we present the performance comparison between SDA and DA implementations of the P2L operator across expansion orders one to ten. As with the P2M case, exact agreement is observed at the constant term, with errors beginning to appear at the first order. These errors remain extremely small: the maximum error recorded is $2.33\times 10^{-12}$, leading to a relative error of only $1.64 \times 10^{-14}$, at order 10, and for most orders the relative error stays well below $1 \times 10^{-14}$, confirming the consistency between the symbolic and numerical approaches. The final three columns of Table~\ref{tab:cl} provide a runtime comparison. Similar to the P2M results, the SDA implementation of the P2L operator significantly outperforms the DA version in computational speed. For example, when the cut-off order $p=9$, the average runtime drops from 12105.59 ns in the DA version to just 237.54 ns in SDA, corresponding to only 1.96\% of the DA runtime. Across all tested orders, the SDA operator runs in under 8\% of the time required by the DA operator, and for orders greater than five it runs in under 3\%.  These findings reinforce that the symbolic expressions generated via SDA not only preserve numerical accuracy but also enable substantial acceleration when evaluating higher-order expansions in practice.

\begin{table}[htbp]
	\centering
	\caption{Verification and performance comparison of the P2M operator using SDA and DA methods}
	\label{tab:cm}
\begin{tabular}{rrrrrrc}
	\toprule
	Order & Value & Error & Rel. Error & DA (ns) & SDA (ns) & SDA/DA (\%) \\
	\midrule
	2 & $\mathtt{-2.12e+00}$ & $\mathtt{4.44e-16}$ & $\mathtt{2.10e-16}$ & 350.80 & 26.72 & 7.62 \\
	3 & $\mathtt{7.93e+00}$ & $\mathtt{2.66e-15}$ & $\mathtt{3.36e-16}$ & 477.93 & 27.54 & 5.76 \\
	4 & $\mathtt{-1.67e+01}$ & $\mathtt{1.07e-14}$ & $\mathtt{6.39e-16}$ & 885.14 & 34.58 & 3.91 \\
	5 & $\mathtt{-2.02e+00}$ & $\mathtt{7.46e-14}$ & $\mathtt{3.69e-14}$ & 1329.49 & 43.66 & 3.28 \\
	6 & $\mathtt{-1.28e+02}$ & $\mathtt{2.84e-13}$ & $\mathtt{2.22e-15}$ & 2646.77 & 68.10 & 2.57 \\
	7 & $\mathtt{1.05e+03}$ & $\mathtt{2.50e-12}$ & $\mathtt{2.38e-15}$ & 4362.84 & 88.70 & 2.03 \\
	8 & $\mathtt{-8.88e+02}$ & $\mathtt{5.57e-12}$ & $\mathtt{6.28e-15}$ & 7008.28 & 120.55 & 1.72 \\
	9 & $\mathtt{7.16e+03}$ & $\mathtt{4.00e-11}$ & $\mathtt{5.59e-15}$ & 11682.09 & 188.23 & 1.61 \\
	10 & $\mathtt{-3.30e+04}$ & $\mathtt{2.62e-10}$ & $\mathtt{7.95e-15}$ & 18515.44 & 441.34 & 2.38 \\
	\bottomrule
\end{tabular}
\end{table}

\begin{table}[htbp]
	\centering
	\caption{Verification and performance comparison of the P2L operator using SDA and DA methods}
	\label{tab:cl}
\begin{tabular}{rrrrrrc}
	\toprule
	Order & Value & Error & Rel. Error & DA (ns) & SDA (ns) & SDA/DA (\%) \\
	\midrule
	1 & $\mathtt{4.94e-01}$ & $\mathtt{2.22e-16}$ & $\mathtt{4.50e-16}$ &  &  &  \\
	2 & $\mathtt{-8.67e-01}$ & $\mathtt{5.55e-16}$ & $\mathtt{6.40e-16}$ & 351.57 & 27.96 & 7.95 \\
	3 & $\mathtt{2.42e+00}$ & $\mathtt{1.78e-15}$ & $\mathtt{7.33e-16}$ & 569.10 & 32.33 & 5.68 \\
	4 & $\mathtt{3.37e+00}$ & $\mathtt{4.88e-15}$ & $\mathtt{1.45e-15}$ & 816.65 & 35.77 & 4.38 \\
	5 & $\mathtt{-8.64e+00}$ & $\mathtt{1.24e-14}$ & $\mathtt{1.44e-15}$ & 1522.79 & 53.51 & 3.51 \\
	6 & $\mathtt{2.15e+01}$ & $\mathtt{3.55e-14}$ & $\mathtt{1.65e-15}$ & 2481.64 & 73.60 & 2.97 \\
	7 & $\mathtt{-4.68e+01}$ & $\mathtt{1.14e-13}$ & $\mathtt{2.43e-15}$ & 4267.13 & 103.90 & 2.43 \\
	8 & $\mathtt{-6.44e+01}$ & $\mathtt{1.99e-13}$ & $\mathtt{3.09e-15}$ & 7184.29 & 144.45 & 2.01 \\
	9 & $\mathtt{1.73e+02}$ & $\mathtt{8.24e-13}$ & $\mathtt{4.76e-15}$ & 12105.59 & 237.54 & 1.96 \\
	10 & $\mathtt{1.43e+02}$ & $\mathtt{2.33e-12}$ & $\mathtt{1.64e-14}$ & 18630.42 & 508.02 & 2.73 \\
	\bottomrule
\end{tabular}
\end{table}

Due to the analogous mathematical formulation of the P2M and P2L operators, their performance is also closely aligned. We observe it in  both the DA and SDA implementations, as expected.

\subsection{M2P and L2P}

\begin{table}[htbp]
	\centering
	\caption{Verification and performance comparison of the M2P operator using SDA and DA methods}
	\label{tab:mc}
\begin{tabular}{rrrrrrS}
	\toprule
	Order & Value & Error & Rel. Error & DA (ns) & SDA (ns) & {SDA/DA (\%)} \\
	\midrule
	2 & $\mathtt{3.26e-01}$ & $\mathtt{2.22e-16}$ & $\mathtt{6.81e-16}$ & 3389.37 & 124.19 & 3.66 \\
	3 & $\mathtt{3.60e-01}$ & $\mathtt{2.78e-16}$ & $\mathtt{7.70e-16}$ & 2048.81 & 100.92 & 4.93 \\
	4 & $\mathtt{3.12e-01}$ & $\mathtt{4.44e-16}$ & $\mathtt{1.42e-15}$ & 2549.71 & 127.28 & 4.99 \\
	5 & $\mathtt{3.03e-01}$ & $\mathtt{5.55e-16}$ & $\mathtt{1.83e-15}$ & 5028.13 & 333.71 & 6.64 \\
	6 & $\mathtt{3.34e-01}$ & $\mathtt{6.11e-16}$ & $\mathtt{1.83e-15}$ & 6061.69 & 461.97 & 7.62 \\
	7 & $\mathtt{3.34e-01}$ & $\mathtt{7.77e-16}$ & $\mathtt{2.33e-15}$ & 6098.66 & 554.02 & 9.08 \\
	8 & $\mathtt{3.42e-01}$ & $\mathtt{7.77e-16}$ & $\mathtt{2.28e-15}$ & 5618.32 & 655.49 & 11.67 \\
	9 & $\mathtt{3.03e-01}$ & $\mathtt{1.17e-15}$ & $\mathtt{3.84e-15}$ & 5014.05 & 630.99 & 12.58 \\
	10 & $\mathtt{3.54e-01}$ & $\mathtt{1.11e-15}$ & $\mathtt{3.13e-15}$ & 4911.70 & 679.82 & 13.84 \\
	\bottomrule
\end{tabular}

\end{table}
\begin{table}[htbp]
	\centering
	\caption{Verification and performance comparison of the L2P operator using SDA and DA methods}
	\label{tab:lc}
\begin{tabular}{rrrrrrS}
	\toprule
	Order & Value & Error & Rel. Error & DA (ns) & SDA (ns) & {SDA/DA (\%)} \\
	\midrule
	2 & $\mathtt{5.08e-01}$ & $\mathtt{2.22e-16}$ & $\mathtt{4.38e-16}$ & 2364.54 & 73.05 & 3.09 \\
	3 & $\mathtt{5.14e-01}$ & $\mathtt{4.44e-16}$ & $\mathtt{8.63e-16}$ & 2009.84 & 74.42 & 3.70 \\
	4 & $\mathtt{5.22e-01}$ & $\mathtt{4.44e-16}$ & $\mathtt{8.52e-16}$ & 1919.79 & 87.50 & 4.56 \\
	5 & $\mathtt{5.22e-01}$ & $\mathtt{6.66e-16}$ & $\mathtt{1.28e-15}$ & 2451.84 & 149.66 & 6.10 \\
	6 & $\mathtt{5.10e-01}$ & $\mathtt{6.66e-16}$ & $\mathtt{1.31e-15}$ & 3355.84 & 221.45 & 6.60 \\
	7 & $\mathtt{5.08e-01}$ & $\mathtt{7.77e-16}$ & $\mathtt{1.53e-15}$ & 3148.84 & 251.20 & 7.98 \\
	8 & $\mathtt{5.22e-01}$ & $\mathtt{1.11e-15}$ & $\mathtt{2.13e-15}$ & 3445.13 & 311.48 & 9.04 \\
	9 & $\mathtt{5.22e-01}$ & $\mathtt{1.33e-15}$ & $\mathtt{2.55e-15}$ & 4879.53 & 667.32 & 13.68 \\
	10 & $\mathtt{5.06e-01}$ & $\mathtt{1.33e-15}$ & $\mathtt{2.63e-15}$ & 6896.87 & 947.86 & 13.74 \\
	\bottomrule
\end{tabular}
\end{table}
The computational procedures for the M2P and L2P operators are identical. Both take as input a full-rank 3D DA vector and three double-precision floating-point numbers. These three numbers are used to assign values to the three DA bases in the vector, and the result is a single double-precision number representing the potential at the particle’s position. The computation involves evaluating a large number of monomials and summing their contributions, which is performed by a shared code for both operators.

The only difference between M2P and L2P lies in the interpretation of the DA bases. In L2P, the base $\mathbf{d}^\prime$, as defined in Eq.~(\ref{eq:P2L}), represents the position of the target particle within the target box. In the case of M2P, even when the position of the target particle relative to the center of the source box is known, an additional translation is still required to compute $\mathbf{d}$, as shown in Eq.~(\ref{eq:P2M}). An extra multiplication of two numbers is needed to obtain $\phi$ from $\phi_\mathrm{M}$ but its effect on the runtime can be ignored. Among all operators in the DA-based FMM, M2P and L2P are the least computationally intensive.

Tables \ref{tab:mc} and \ref{tab:lc} present a comparison between the SDA and DA implementations of the M2P and L2P operators, respectively. Since these two operators share the same evaluation routine, their accuracy and performance are expected to follow similar trends. As shown in Table~\ref{tab:mc}, the maximum absolute and relative errors across orders 2 to 10 remain very low, typically on the order of $10^{-15}$ or lower, indicating excellent agreement between the SDA- and DA-based evaluations. The same conclusion holds for L2P, as shown in Table~\ref{tab:lc}. The relative errors in both tables do not exceed $2.63 \times 10^{-15}$, and most are well below that threshold. In terms of runtime, however, the SDA implementations are consistently and significantly faster than their DA counterparts. For M2P (Table~\ref{tab:mc}), the average runtime of SDA stays between 3.66\% and 13.84\% of the DA version. For L2P (Table~\ref{tab:lc}), the corresponding range is 3.09\% to 13.74\%. Unlike the observations from from the P2M and P2L benchmarks.  The efficiency gains are more pronounced for the lower orders than the higher orders.  Overall, these results confirm the numerical fidelity and the substantial improvements in computational efficiency for the SDA M2P and SDA L2P operators.

\subsection{M2M and L2L}

The M2M and L2L operators are appreciably more demanding than the four operators discussed earlier. Their evaluation involves the composition of two DA vectors: the original multipole or local expansion, and a DA map that converts the DA bases from the old to the new coordinate frame. Composing these two vectors—effectively substituting one DA vector into another—produces a new DA vector. For L2L, the new DA vector represents the expansion in the shifted coordinate system. But the M2M operator performs an extra DA-vector multiplication on top of it to obtain the new expansion, as expressed in Eq.~(\ref{eq:m2m}).

The M2M operator involves composing two 3-D full-rank DA vectors, a step that is normally computationally heavy. However, because in FMM we usually decompose the space under study into cubic boxes with a hierarchical tree structure, we can exploit geometric symmetry to collapse the map $M_1$ in Eq.~(\ref{eq:m2m}) from a three-variable map to a one-variable map. Specifically,  $M_1$ depends on the shift vector $\mathbf{r}_0$ that points from the child-box center to the parent-box center. Because every box is a cube, the three Cartesian components of $\mathbf{r}_0$ always have the same magnitude; only their signs differ, determined by the relative octant in which the child box lies. Consequently, 
$M_1$
can be rewritten as eight maps of a single scalar variable - the common magnitude of those components. A simple function that handles the sign pattern decides which map to choose for a specific translation. This dimensional reduction greatly simplifies the resulting coefficients produced by the composition and, in turn, leads to a substantial speed-up in the generated evaluation code.

In contrast, the computation of the L2L operator is significantly simpler. In Eq.~(\ref{eq:l2l}), the second DA vector involved in the composition, $M_3$, represents a linear shift—that is, it has rank 1 and contains no second- or higher-order terms. In principle, $M_3$ could also be reduced to eight single-variable functions, corresponding to the eight possible relative positions between a parent box and its children. However, we did not apply this optimization, as the simplicity of $M_3$ already keeps the composition lightweight and efficient in practice.

Tables~\ref{tab:mm} and~\ref{tab:ll} present the accuracy and runtime comparison of the M2M and L2L operators computed using SDA and DA. 

For M2M (Table~\ref{tab:mm}), errors start to appear at the second order. The maximum absolute error and the corresponding relative errors remain small, generally below $1\times 10^{-15}$, indicating excellent numerical agreement. Despite the increased computational complexity due to the composition of two  DA vectors, the SDA implementation remains efficient. Across all tested orders, the SDA runtime is consistently under 2.5\% of the DA runtime, with some cases as low as 1.98\%. This confirms that the symbolic simplification strategy—especially the reduction of $M_1$ to single-variable functions—yields not only simpler expressions but also highly efficient code.

For L2L (Table~\ref{tab:ll}), errors start to show from the constant term and similar accuracy is observed, with relative errors mostly on the order of $1\times 10^{-15}$ or smaller. At order 10, the DA and SDA outputs are exactly the same because a linear shift only leads to changes in lower order terms and hence the terms in the highest order remains exactly the same.  The SDA runtime ranges from 4.33\% to 7.02\% of the DA version for orders 2 through 10.  While the performance gain in L2L is less dramatic than in M2M—primarily due to the simpler structure of the linear shift map $M_3$—the SDA implementation still achieves a consistent 5- to 20-fold speedup.

\begin{table}[htbp]
	\centering
	\caption{Verification and performance comparison of the M2M operator using SDA and DA methods}
	\label{tab:mm}
\begin{tabular}{rrrrrrc}
	\toprule
	Order & Value & Error & Rel. Error & DA ($\mu$s) & SDA ($\mu$s) & SDA/DA (\%) \\
	\midrule
	2 & $\mathtt{-2.44e+00}$ & $\mathtt{8.88e-16}$ & $\mathtt{3.64e-16}$ & 1.61 & 0.04 & 2.49 \\
	3 & $\mathtt{1.02e+01}$ & $\mathtt{5.33e-15}$ & $\mathtt{5.22e-16}$ & 2.81 & 0.06 & 2.02 \\
	4 & $\mathtt{-1.83e+01}$ & $\mathtt{1.07e-14}$ & $\mathtt{5.81e-16}$ & 5.75 & 0.11 & 1.98 \\
	5 & $\mathtt{-9.44e+01}$ & $\mathtt{5.68e-14}$ & $\mathtt{6.02e-16}$ & 12.64 & 0.30 & 2.41 \\
	6 & $\mathtt{3.50e+02}$ & $\mathtt{2.84e-13}$ & $\mathtt{8.11e-16}$ & 31.19 & 0.73 & 2.33 \\
	7 & $\mathtt{1.09e+03}$ & $\mathtt{6.82e-13}$ & $\mathtt{6.26e-16}$ & 73.74 & 1.60 & 2.17 \\
	8 & $\mathtt{2.53e+03}$ & $\mathtt{3.18e-12}$ & $\mathtt{1.26e-15}$ & 164.38 & 3.61 & 2.20 \\
	9 & $\mathtt{-1.19e+04}$ & $\mathtt{1.27e-11}$ & $\mathtt{1.07e-15}$ & 335.25 & 6.62 & 1.98 \\
	10 & $\mathtt{4.07e+04}$ & $\mathtt{4.37e-11}$ & $\mathtt{1.07e-15}$ & 698.41 & 15.47 & 2.22 \\
	\bottomrule
\end{tabular}
\end{table}

\begin{table}[htbp]
	\centering
	\caption{Verification and performance comparison of the L2L operator using SDA and DA methods}
	\label{tab:ll}
\begin{tabular}{rrrrrrc}
	\toprule
	Order & Value & Error & Rel. Error & DA ($\mu$s) & SDA ($\mu$s) & SDA/DA (\%) \\
	\midrule
	0 & $\mathtt{1.09e+00}$ & $\mathtt{2.89e-15}$ & $\mathtt{2.65e-15}$ &  &  &  \\
	1 & $\mathtt{1.23e+00}$ & $\mathtt{3.33e-15}$ & $\mathtt{2.71e-15}$ &  &  &  \\
	2 & $\mathtt{-5.73e+00}$ & $\mathtt{9.77e-15}$ & $\mathtt{1.71e-15}$ & 0.93 & 0.05 & 5.05 \\
	3 & $\mathtt{1.96e+01}$ & $\mathtt{1.42e-14}$ & $\mathtt{7.25e-16}$ & 1.47 & 0.06 & 4.37 \\
	4 & $\mathtt{-5.12e+01}$ & $\mathtt{3.55e-14}$ & $\mathtt{6.94e-16}$ & 2.51 & 0.11 & 4.33 \\
	5 & $\mathtt{2.04e+02}$ & $\mathtt{8.53e-14}$ & $\mathtt{4.17e-16}$ & 4.10 & 0.22 & 5.40 \\
	6 & $\mathtt{4.94e+02}$ & $\mathtt{1.14e-13}$ & $\mathtt{2.30e-16}$ & 7.74 & 0.54 & 7.02 \\
	7 & $\mathtt{3.84e+02}$ & $\mathtt{2.27e-13}$ & $\mathtt{5.92e-16}$ & 14.02 & 0.94 & 6.74 \\
	8 & $\mathtt{1.17e+03}$ & $\mathtt{2.27e-13}$ & $\mathtt{1.95e-16}$ & 27.51 & 1.66 & 6.03 \\
	9 & $\mathtt{7.90e+02}$ & $\mathtt{1.14e-13}$ & $\mathtt{1.44e-16}$ & 50.03 & 2.50 & 4.99 \\
	10 & &  &  & 91.08 & 3.47 & 3.81 \\
	\bottomrule
\end{tabular}
\end{table}

\begin{table}[htbp]
	\centering
	\caption{Verification and performance comparison of the M2L operator using SDA and DA methods}
	\label{tab:ml}
	\begin{tabular}{rrrrrrc}
		\toprule
		Order & Value & Error & Rel. Error & DA ($\mu$s) & SDA ($\mu$s) & SDA/DA (\%) \\
		\midrule
		0 & $\mathtt{1.62e-01}$ & $\mathtt{1.39e-16}$ & $\mathtt{8.58e-16}$ &  &  &  \\
		1 & $\mathtt{-1.96e-02}$ & $\mathtt{1.39e-17}$ & $\mathtt{7.07e-16}$ &  &  &  \\
		2 & $\mathtt{5.75e-03}$ & $\mathtt{4.34e-18}$ & $\mathtt{7.55e-16}$ & 4.27 & 0.17 & 4.00 \\
		3 & $\mathtt{-3.78e-03}$ & $\mathtt{3.04e-18}$ & $\mathtt{8.02e-16}$ & 6.55 & 0.23 & 3.48 \\
		4 & $\mathtt{-8.27e-04}$ & $\mathtt{7.59e-19}$ & $\mathtt{9.18e-16}$ & 13.05 & 0.56 & 4.29 \\
		5 & $\mathtt{-3.08e-04}$ & $\mathtt{5.42e-19}$ & $\mathtt{1.76e-15}$ & 25.60 & 1.24 & 4.85 \\
		6 & $\mathtt{1.48e-04}$ & $\mathtt{2.71e-19}$ & $\mathtt{1.83e-15}$ & 55.55 & 2.63 & 4.74 \\
		7 & $\mathtt{3.14e-05}$ & $\mathtt{1.36e-19}$ & $\mathtt{4.31e-15}$ & 110.53 & 5.83 & 5.28 \\
		8 & $\mathtt{-1.65e-05}$ & $\mathtt{9.15e-20}$ & $\mathtt{5.55e-15}$ & 213.15 & 11.44 & 5.37 \\
		9 & $\mathtt{2.23e-06}$ & $\mathtt{4.57e-20}$ & $\mathtt{2.05e-14}$ & 396.04 & 19.88 & 5.02 \\
		10 & $\mathtt{-2.35e-06}$ & $\mathtt{2.67e-20}$ & $\mathtt{1.14e-14}$ & 741.93 & 35.83 & 4.83 \\
		\bottomrule
	\end{tabular}
\end{table}

\subsection{M2L}
It is well-known that the M2L process is the most time-consuming component in an FMM algorithm—not only because each individual M2L operation is relatively expensive, but also because it occurs far more frequently than M2M or L2L operations. In 3D FMM, a box has at most 8 child boxes, requiring up to 8 M2M or L2L operations. In contrast, it can have as many as 189 boxes in its interaction list, each triggering a separate M2L translation. This large disparity in invocation frequency means that the overall performance of the FMM algorithm is heavily influenced by the efficiency of the M2L operator. Therefore, improving the speed of the M2L computation is critical to enhancing the performance of the entire DA-FMM algorithm.

From a computational standpoint, the M2L kernel is almost identical to M2M.
The procedure can be summarized as follows: We first build the transfer map $M_2$, a full-rank 3D DA vector that converts the DA bases from the source‐box frame to the target‐box frame. During this step we also obtain an auxiliary full-rank vector $R$. Then we compose the source-box multipole expansion (another full-rank DA vector) with $M_2$.
This composition of two 3D full-rank DA vectors is by far the most expensive part of the computation. Finally, we multiply the composition result by the square root of $R$.
For M2M we could exploit the symmetry of cubic boxes to collapse the base-transfer map $M_1$ to single-variable functions, dramatically reducing both the algebraic complexity and the runtime.
For M2L, however, the geometric relationship between the source and target boxes is more complicated, so the same symmetry trick is not available. 
We experimented with direct code generation from the 3D SDA composition:
At very low orders ($p$ = 2 or 3), the SDA version cut the runtime of a single M2L call roughly in half. Beyond order 5, the expressions ballooned in size, and the SDA implementation actually ran slower than the numerical DA version. Hence a naive SDA-based code generation does not meet our performance goals.

The key observation is that the transfer map $M$ depends only on the displacement vector
$\mathbf{r}=(x,y,z)$ that connects the centers of the source and target boxes. In the source-box frame, this vector originates at the origin. If we rotate the coordinate system such that $\mathbf{r}$ lies along a coordinate axis, $M$ becomes a single-variable function \cite{abeyratne2013optimization}. This can be achieved in two steps. First, we rotate the coordinate system around the $x$-axis by an angle $\theta=\arctan(z/y)$, which brings the target box center into the 
$x-y$ plane. Then, we perform a second rotation around the $z$-axis by an angle $\phi=\arctan(y^2+z^2/x)$, which aligns the target box center with the new $x$-axis. After these rotations, the displacement vector becomes $\mathbf{r}^\prime=(r_x,0,0)$, where $r_x =\sqrt{x^2+y^2+z^2}$, since the length of $\mathbf{r}$ is invariant under rotation. $M$ now depends on a single scalar $r_x$. The entire M2L computation can then be carried out in the rotated frame, and the result transformed back to the original coordinate system using the inverse rotation. Both the forward and the inverse rotations can be represented by the respective $3\times 3$ 
orthogonal matrix, or equivalently, by the DA map containing only first-order terms. Therefore, rotating a full-rank DA vector to the new frame or back requires only the composition with a rank-1 DA vector. This effectively replaces one expensive 3D composition with three much simpler compositions, resulting in significantly simplified expressions in the SDA calculation and a substantial improvement in computational efficiency of the generated code.

Table \ref{tab:ml} reports the accuracy and runtime of the M2L operator up to order 10 when both the DA and SDA implementation employ the same axis-rotation trick to align the displacement vector $r$ with a coordinate axis. Even with this optimisation applied to both codes, SDA remains decisively faster while preserving near machine–precision agreement: the largest absolute errors and the respective relative errors stay in the $10^{-16}$ to $10^{-14}$ range throughout.

In terms of performance, the SDA-based M2L operator consistently outperforms the DA version. For orders 2 through 10, the SDA runtime remains between 3.48\% and 5.37\% of the DA runtime, achieving roughly a 20-fold speedup in most cases.  Comparing these figures with those in Table \ref{tab:mm} for the M2M operator, we observe that the DA M2L requires slightly more time than DA M2M and the SDA M2L does not achieve the 50-fold speedup as the SDA M2M does. Nevertheless, the 20-fold speedup remains satisfactory. 
Overall, applying the rotation-based simplification in combination with SDA yields generated code that runs markedly faster than its numerical DA counterpart while retaining its high numerical fidelity. It is important to note that this rotation trick does not depend on the specific form of the kernel function and can thus be employed as a general strategy to simplify the DA M2L operator for any non-oscillatory kernel.

\subsection{Numerical property of the enhanced DA-FMM}
We have demonstrated that for each individual operator, high-efficiency code can be generated based on the SDA formulation, achieving a 20- to 50-fold speedup compared to the corresponding DA implementations. In the this section, we will evaluate the overall numerical performance of the SDA-FMM and benchmark it against two state-of-the-art FMM implementations: the traceless Cartesian tensor-based FMM  \cite{huang2018improve} and pyfmmlib \cite{pyfmmlib2025}. To construct the SDA-FMM \cite{SDAFMMCode}, we adopt the FMM framework from the open-source traceless Cartesian tensor-based code \cite{TracelessCartesianTensorFMM} and systematically replace all kernel operators with their SDA counterparts. In all numerical experiments presented below, particles are sampled from a three-dimensional Gaussian distribution with unit standard deviation.

The error of the DA-based FMM is  governed by the truncation order of the DA vectors. Higher orders yield smaller errors. To examine the convergence behaviour, we computed the Coulomb potential between 1,000,000 particles with the SDA-FMM code while varying the expansion order from two to ten. The results were compared with with the analytical solution, and the relative root-mean-square error was evaluated as
\begin{equation}
	\sigma = \sqrt{\frac{\sum_{i=1}^{N}(\phi(\mathbf{r}_i)-\phi_0(\mathbf{r}_i))^2}{\sum_{i=1}^{N}\phi_0^2(\mathbf{r}_i)}},
\end{equation}
where $\phi$ is the potential calculated by the SDA-FMM code,  $\phi_0$ is the exact value of the potential, $N$ is the number of particles, and $\mathbf{r}_i$ is the position of the $i$-th particle. As shown in Table~\ref{tab:error}, the error decreases monotonically from 
$7.94\times 10^{-4}$ at order 2 to  $2.95\times 10^{-8}$  at order 10. On average, increasing the expansion order by two reduces the error by roughly one order of magnitude. 

\begin{figure}[htbp]
	\centering
	
	\begin{subfigure}[b]{0.48\textwidth}
		\includegraphics[width=\linewidth]{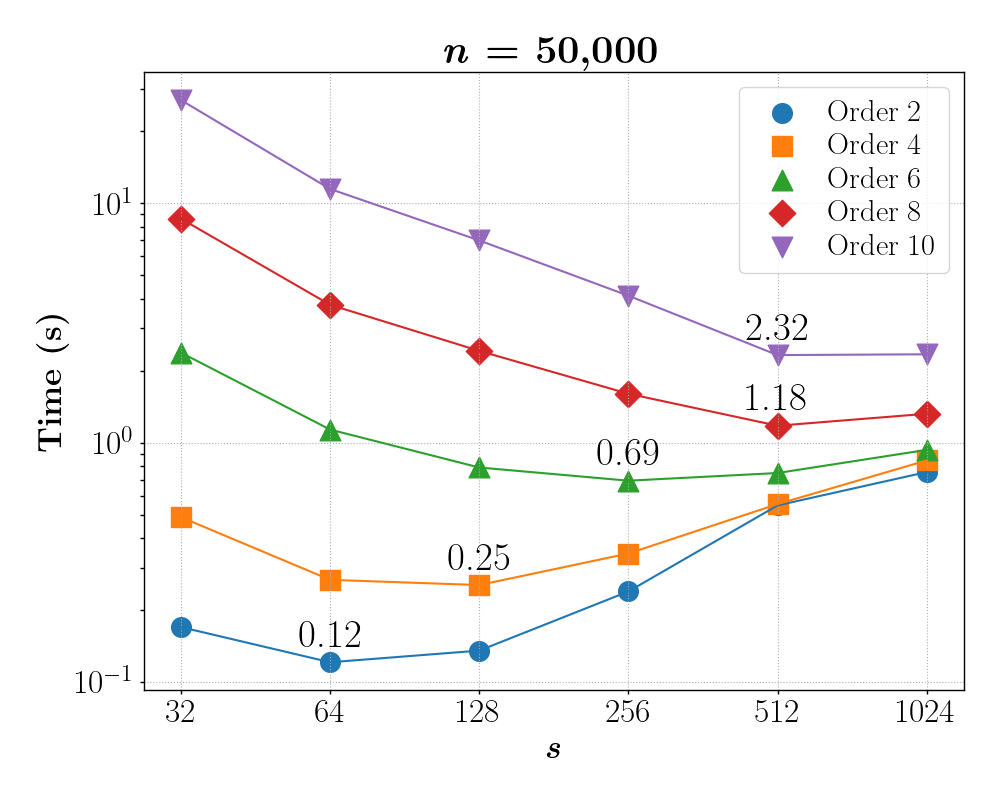}
	\end{subfigure}
	\hfill
	\begin{subfigure}[b]{0.48\textwidth}
		\includegraphics[width=\linewidth]{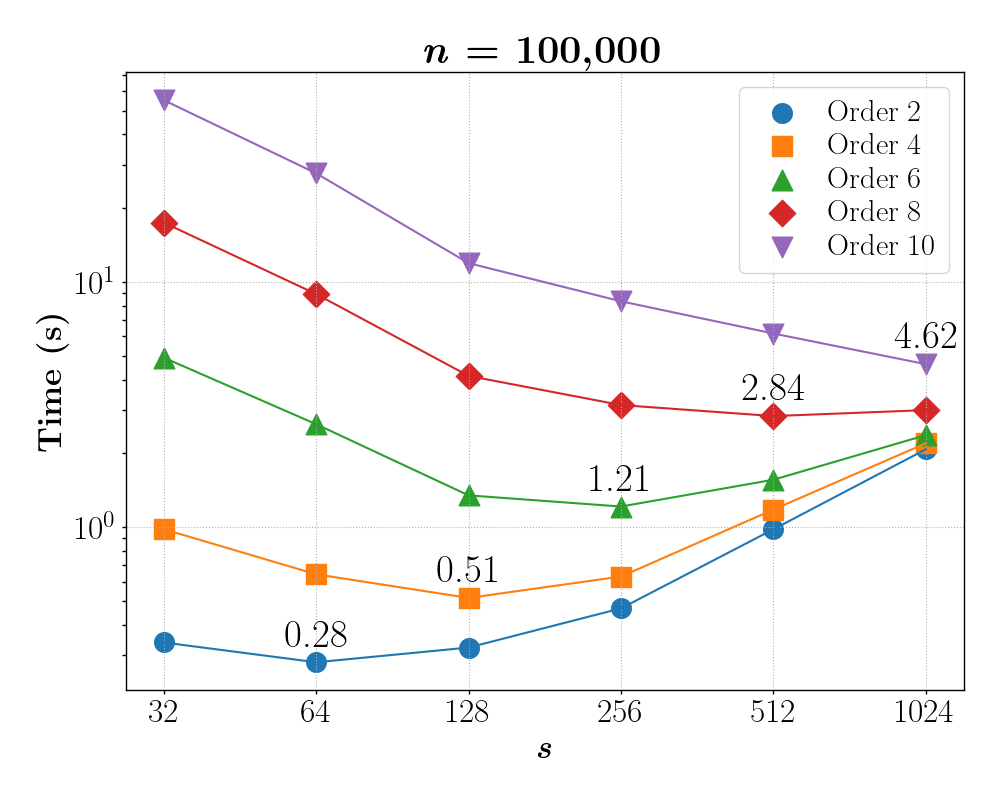}
	\end{subfigure}
	
	\vskip\baselineskip
	
	\begin{subfigure}[b]{0.48\textwidth}
		\includegraphics[width=\linewidth]{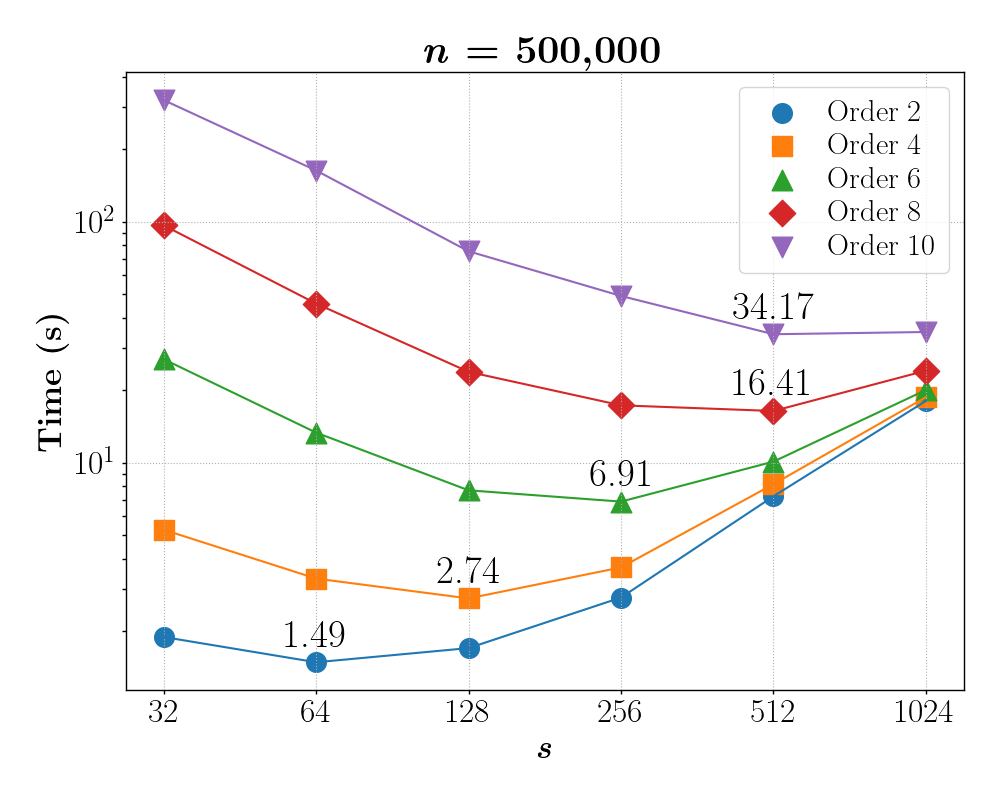}
	\end{subfigure}
	\hfill
	\begin{subfigure}[b]{0.48\textwidth}
		\includegraphics[width=\linewidth]{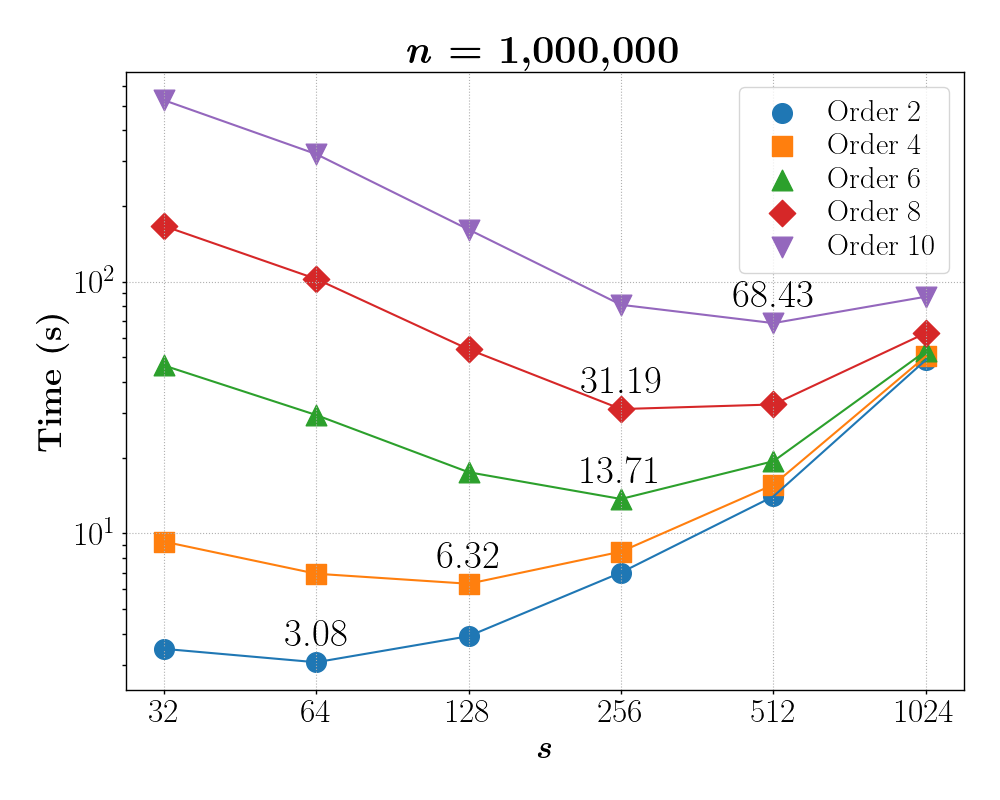}
	\end{subfigure}
	
	\vskip\baselineskip
	
	\begin{subfigure}[b]{0.48\textwidth}
		\includegraphics[width=\linewidth]{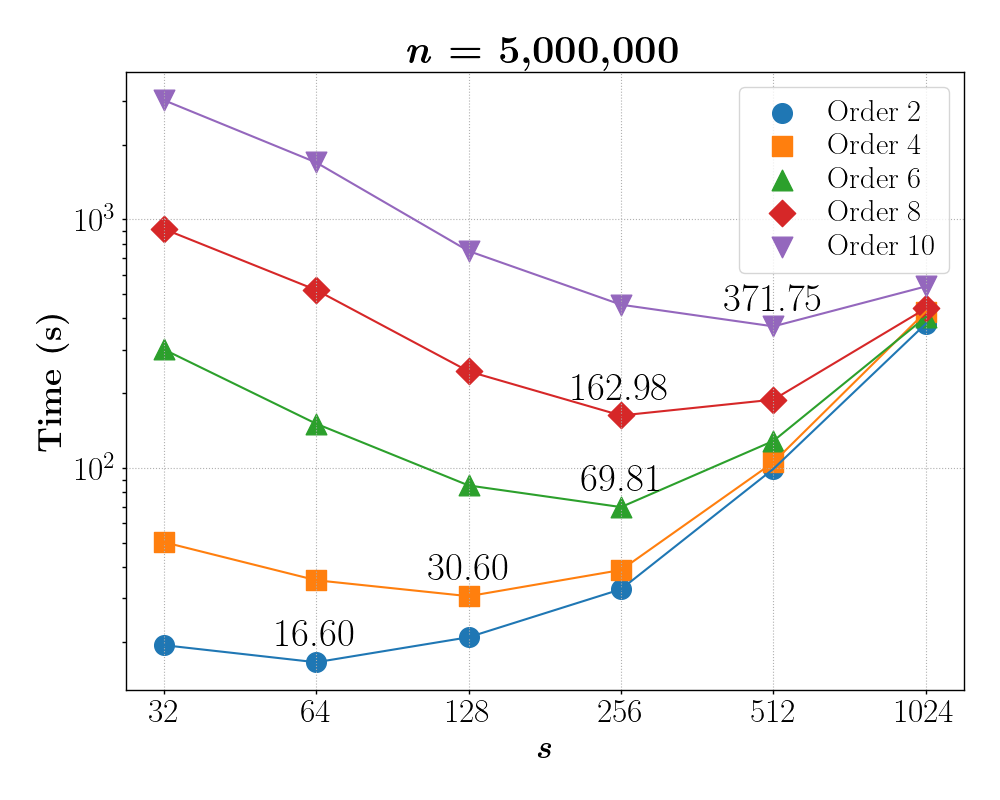}
	\end{subfigure}
	\hfill
	\begin{subfigure}[b]{0.5\textwidth}
		\includegraphics[width=\linewidth]{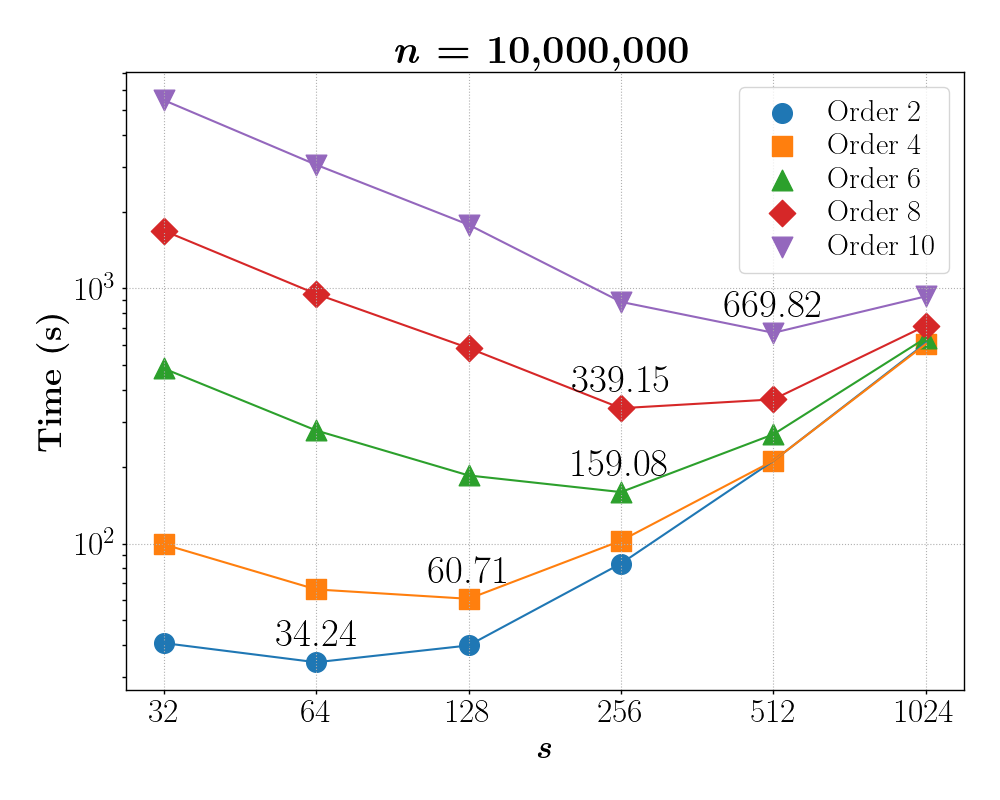}
	\end{subfigure}
	
	\caption{Computation time for different particle number $n$, expansion order $p$, and leaf size $s$.}
	\label{fig:best_s}
\end{figure}

\begin{table}[htbp]
	\centering
	\caption{Relative error for different orders}
	\label{tab:error}
	\begin{tabular}{clllll}
		\toprule
		order & 2 & 4 & 6 & 8 & 10 \\
		error & $\mathtt{7.94e-04}$ &$\mathtt{6.30e-05}$  &$\mathtt{3.76e-06}$ &$\mathtt{1.58e-07}$ & $\mathtt{2.95e-08}$ \\
		\bottomrule
	\end{tabular}	
\end{table}

In the FMM algorithm, the maximum number of particles allowed in a childless box is controlled by a parameter $s$, which we refer to as the leaf size. Any box with more than $s$ particles will be subdivided into smaller child boxes. The choice of $s$ has a substantial impact on the runtime of the FMM algorithm. To identify the optimal leaf size that yields the shortest runtime, we conducted a series of experiments across different particle numbers and truncation orders. The results are shown in Fig.~\ref{fig:best_s}. Each subplot corresponds to a different number of particles, ranging from 50,000 to 10,000,000. Within each subplot, the expansion order varies from two to ten, and $s$ is swept from 32 to 1024. For each order, the data point with the minimum runtime is annotated with its corresponding value. 

We summarize the optimal leaf size for different particle numbers and truncation orders in table~\ref{tab:s}. The data reveal a clear correlation between the truncation order and the optimal leaf size: higher truncation orders favor larger $s$. Only for order $p=8$ do we see the optimal $s$ decrease as the particle number grows:  $s=512$ gives the shortest runtimes for  500,000 or more particles while $s=256$ becomes optimal for 1,000,000 or more particles. For order $p=10$, all test cases prefer $s=512$ except a single instance with 100,000 particles, which selects $s=1024$. When the truncation order is six or lower, the optimal $s$ is essentially insensitive to the particle number.  These observations offer practical guidance for choosing the leaf size parameter in SDA–FMM simulations across a wide range of problem sizes and accuracy requirements.

\begin{table}[htbp]
	\centering
	\caption{Optimal $s$ for shortest computation time  with various particle numbers and orders of the expansions}
	\label{tab:s}
	\begin{tabular}{cccccccc}
		\toprule
		\multicolumn{1}{c}{Order} & \multicolumn{1}{c}{10,000} & \multicolumn{1}{c}{50,000}
		  & \multicolumn{1}{c}{100,000}  & \multicolumn{1}{c}{500,000} 
		   & \multicolumn{1}{c}{1,000,000}  & \multicolumn{1}{c}{5,000,000}
		    & \multicolumn{1}{c}{10,000,000}\\
		%		DA & SDA & Orders \\
		\hline
		2   &  64  & 64 & 64 & 64 & 64 & 64  &  64 \\
		4   &   128 & 128 & 128 & 128 & 128 & 128  & 128  \\
		6   &  512  & 256 & 256 & 256  &  256 & 256 &  256 \\
		8   &   512 & 512 & 512  & 512 & 256 & 256 &256 \\
		10   &  512  & 512  &  1024 & 512 & 512 & 512 &  512 \\		
		\bottomrule
	\end{tabular}
\end{table}

\begin{figure}[htbp]
	\centering
	\includegraphics[width=0.6\textwidth]{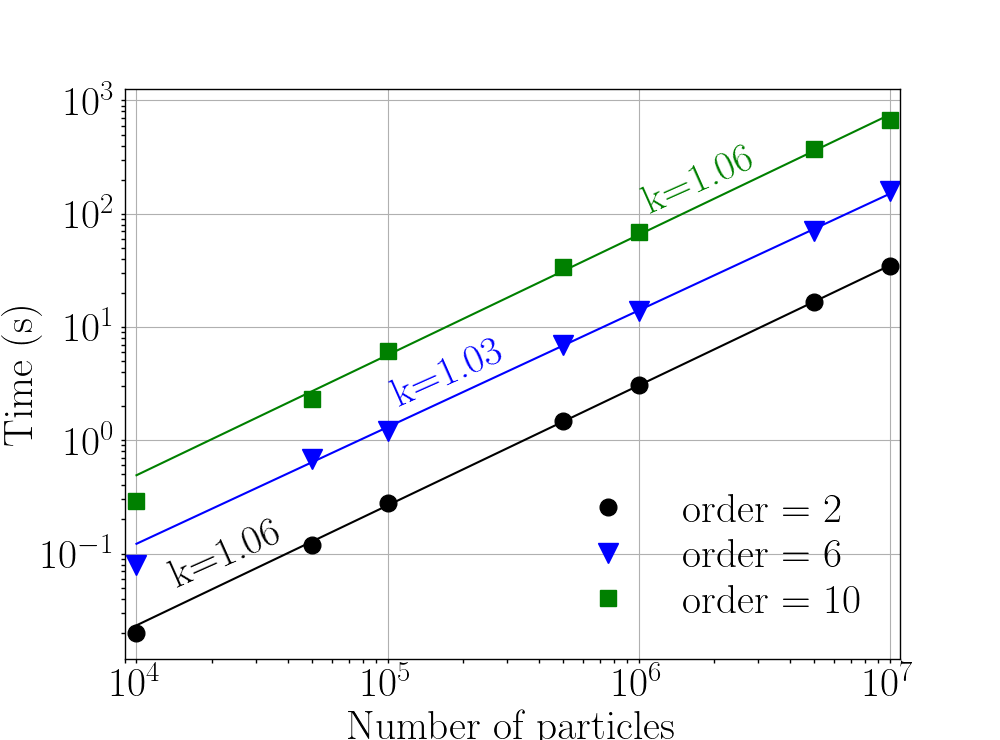}
	\caption{Linear scaling with particle number.}
	\label{fig:linear}
\end{figure}

The linear scaling of computation time with respect to the number of particles is well-known for algorithms in the FMM family. To verify this property for SDA-FMM, we computed the Coulomb potential for particle counts ranging from 10,000 to 10,000,000, using truncation orders of two, six, and ten, and setting $s$ to its optimal value in each case. The results are shown in Fig.~\ref{fig:linear}, where both the computation time and the number of particles are plotted on logarithmic scales. For each truncation order, the data points can be fitted by a straight line, and the corresponding slope is annotated. All slopes are very close to one, confirming that the runtime of the SDA-FMM grows linearly with the number of particles. This result validates that the SDA-based implementation preserves the hallmark scalability of the FMM framework.

\begin{table}[htbp]
	\centering
	\caption{Time cost and relative error for Coulomb potential between 1,000,000 particles}
	\label{tab:benchmark}
\begin{tabular}{ccS@{\hskip 2em}cccS@{\hskip 2em}cS[table-format=3.0, round-mode=places, round-precision=0]cS}
	\toprule
	\multicolumn{3}{c}{\textbf{pyfmmlib}} & \multicolumn{4}{c}{\textbf{Traceless Cartesian tensor FMM}} & \multicolumn{4}{c}{\textbf{SDA enhanced FMM}} \\
	iprec & $\sigma_\phi$ & {$T_\phi$ (s)} & rank & $s$ & $\sigma_\phi$ & {$T_\phi$ (s)} & rank & $s$ & $\sigma_\phi$ & {$T_\phi$ (s)} \\
	\midrule
	-2 & $\mathtt{8.10e-04}$ & 9.83 & 3 & 128 & $\mathtt{4.13e-04}$ & 6.62 & 3 & 64 & $\mathtt{3.12e-04}$ & 4.70 \\
	-1 & $\mathtt{1.01e-04}$ & 12.35 & 4 & 128 & $\mathtt{1.15e-04}$ & 9.15 & 4 & 128 & $\mathtt{7.03e-05}$ & 6.14 \\
	0 & $\mathtt{1.96e-06}$ & 21.35 & 10 & 256 & $\mathtt{2.25e-06}$ & 43.60 & 7 & 256 & $\mathtt{8.49e-07}$ & 20.86 \\
	\bottomrule
\end{tabular}
\end{table}

Finally, we benchmark the SDA-FMM against both the traceless Cartesian tensor-based FMM and pyfmmlib for Coulomb potential evaluation involving 1,000,000 particles. The traceless Cartesian tensor-based FMM represents multipole and local expansions using Cartesian tensors and leverages the traceless property to reduce redundancy and improve computational efficiency \cite{huang2018improve}. Its accuracy is controlled by the rank of the tensor. Since the leaf size $s$ affects runtime performance, we find the optimal leaf size for all cases of the Cartesian tensor-based FMM used in the benchmark. The SDA-FMM implementation utilizes the same FMM framework in the Cartesian tensor-based FMM code, allowing for a direct comparison of only the kernel implementations. In contrast, pyfmmlib is a Python wrapper around the high-performance Fortran libraries fmmlib2d \cite{fmmlib2d2025} and fmmlib3d \cite{fmmlib3d2025}, which implement the FMM for Laplace and Helmholtz potentials using spherical harmonic functions to represent the multipole and local  expansions. Known for its high efficiency, pyfmmlib is often used as a reference benchmark for FMM performance. In pyfmmlib, the user adjust the error tolerance through the input parameter `iprec`, which ranges from $-2$ to 5; higher values of `iprec` correspond to smaller error. Since the three codes differ in kernel formulation and error control mechanisms, it is difficult to match their output errors exactly. In this benchmark, we ensure that all three implementations achieve comparable accuracy to provide a fair performance comparison. 

The controlling parameters, computation time, and relative errors for all three codes are summarized in Table~\ref{tab:benchmark}.  We first compare the SDA–FMM with the traceless Cartesian tensor–based FMM.  At modest truncation orders ($p=3$ or $4$), SDA–FMM consistently delivers a smaller relative error in a shorter time.  The advantage widens at higher orders: the Cartesian–tensor code attains an error of $2.25\times10^{-6}$ with $p=10$ in 43.60 seconds, whereas SDA–FMM achieves a lower error already at $p=7$ while requiring less than 50\% of the run time (20.86 seconds).  Because the number of tensor coefficients grows rapidly with $p$, it is reasonable to expect SDA–FMM to outperform the Cartesian approach across the full range of practical accuracy. We next compare SDA–FMM with pyfmmlib.  For $\mathrm{iprec}=-2$ and $-1$, pyfmmlib achieves errors on the order of $10^{-4}$.  SDA–FMM reaches still lower errors at the corresponding orders $p=3$ and $p=4$, while consuming roughly half the runtime.  With $\mathrm{iprec}=0$ , pyfmmlib obtains an error of $1.96\times 10^{-6}$ in 21.35 seconds; SDA–FMM attains a 50\% smaller error in 20.86 seconds.  We therefore conclude that, for moderate accuracy requirements with a relative error above $10^{-7}$, SDA–FMM outperforms pyfmmlib.  At very high accuracy, however, the spherical‐harmonic expansion used by pyfmmlib involves fewer terms than a Cartesian expansion and may become faster.  It should be noted that spherical harmonics are available for only a limited number of kernels, whereas both Cartesian tensor–based and DA‐based FMM formulations extend naturally to arbitrary non-oscillatory kernels.

\section{Summary and Broader Impact} \label{sec:summary}

DA offers a convenient and powerful framework for manipulating multivariate Taylor expansions, making it particularly suitable for constructing the FMM based on series expansions. However, traditional DA-based FMM implementations rely on numerical evaluation of DA operations, which is significantly slower than other FMM variants such as those based on Cartesian tensors or spherical harmonics. This performance bottleneck has historically limited the practical applicability of DA in high-performance settings.

The introduction of SDA addresses this issue by enabling the derivation of explicit expressions for all DA-based FMM operators. These expressions can then be used to generate highly efficient, statically optimized code. As demonstrated in our study, the use of SDA leads to substantial computational advantages: we observe a 20- to 50-fold speedup for individual FMM operators when compared to their numerically evaluated DA counterparts.

We further studied the numerical property of the SDA-FMM and benchmarked it against two state-of-the-art alternatives: pyfmmlib, which uses spherical harmonics, and the traceless Cartesian tensor-based FMM. For relative error levels above $10^{-7}$, SDA-FMM consistently outperformed both in terms of runtime while maintaining or exceeding their accuracy. These results demonstrate that the SDA-FMM achieves performance and accuracy that are comparable to state-of-the-art FMM implementations.

Beyond this specific application, the SDA framework provides a general mechanism for resolving the long-standing efficiency concerns associated with numerical DA packages. The methodology of transforming symbolic expressions from SDA into optimized numerical code holds promise not only for FMM but for a broad class of numerical algorithms that rely on Taylor expansion techniques, where the need for both precision and performance is critical.

\section*{Acknowledgement}
This material is based upon work supported by the U.S. Department of Energy, Office of Science, Office of Nuclear Physics under contract DE-AC05-06OR23177.

\appendix

\section{Brief introduction to differential algebra}\label{ap:da}
Consider the vector space of the infinitely differentiable functions
$C^{\infty}(R^{v})$, in which we can define an equivalence relation
``$=_{n}$'' between two functions $a,b\in C^{\infty}(R^{v})$ via
$a=_{n}b$ if $a(0)=b(0)$ and if all the partial derivatives of $a$
and $b$ at 0 agree up to the order $n$. Note that the point 0 is
selected for convenience, and any other point could be chosen as well.
The set of all $b$ that satisfies $b=_{n}a$ is called the equivalence
class of $a,$ which is denoted by $[a]_{n}$. We denote all the equivalence
classes with respect to $=_{n}$ on $C^{\infty}(R^{v})$ as $_{n}D_{v}$.
The addition, scalar multiplication and multiplication on $_{n}D_{v}$
can be defined as Eq.~(\ref{eq:addmlt})

\begin{eqnarray}
[a]_{n}+[b]_{n} & := & [a+b]_{n},\nonumber \\
c\cdot[a]_{n} & := & [c\cdot a]_{n},\label{eq:addmlt}\\
{}[a]_{n}\cdot[b]_{n} & := & [a\cdot b]_{n},\nonumber 
\end{eqnarray}
where $a,b\in{_{n}D_{v}}$ and $c$ is a scalar, so that $_{n}D_{v}$
is an algebra. We can also define the derivation operator $\partial_{v}$
as Eq.~(\ref{eq:dev})
\begin{equation}
\partial_{v}[a]_{n}:=\left[\frac{\partial}{\partial x_{v}}a\right]_{n-1},\label{eq:dev}
\end{equation}
where $x_{v}$ is the $v^{\mathrm{th}}$ variable of the function
$a$. The operator $\partial_{v}$ satisfies 
\begin{equation}
\partial_{v}([a]\cdot[b])=[a]\cdot(\partial_{v}[b])+(\partial_{v}[a])\cdot[b]\label{eq:dev2}
\end{equation}
An algebra with a derivation is called a differential algebra. There
are $v$ special classes $d_{v}=[x_{v}],$ whose elements are all
infinitely small. According to the fixed point theorem\cite{AIEP108book},
the inverse and the roots of any element that is not infinitely small
in $_{n}D_{v}$ exist and can be calculated easily. Further more,
all real power series can be extend to the DA within their radius
of convergence. If a function $a$ in $_{n}D_{v}$ has all the derivatives
$c_{J_{1},...J_{v}}=\partial^{J_{1}+...+J_{v}}a/\partial x_{1}^{J_{1}}\cdot...\cdot\partial x_{v}^{J_{v}},$
then $[a]$ can be written as 
\begin{equation}
[a]=\sum c_{J_{1},...J_{v}}\cdot d_{1}^{J_{1}}\cdot...\cdot d_{v}^{J_{v}}.\label{eq:poly}
\end{equation}
Thus $d_{1}^{J_{1}}\cdot...\cdot d_{v}^{J_{v}}$ is a basis of the
vector space of $_{n}D_{v}$. The Eq.~(\ref{eq:poly}) reminds us
of the Taylor expansion of a function. Actually if we have a function
$f$ in $C^{\infty}(R^{v})$ and $f_{T}$ is its Taylor expansion
up to order $n$, obviously we have $f=_{n}f_{T}$ in $_{n}D_{v}$.
In practice this means we can express $f$ by its Taylor expansion
up to an arbitrary order $n$ as an element in $_{n}D_{v}$, and we
can calculate the derivative classes of $f$ and any other function
that can be derived by applying the elemental operations, divisions,
roots, and power series on $f$. \cite{cpo3ho}

\printbibliography
\end{document}